\documentclass[%reprint,
preprint,
showpacs,%preprintnumbers,
 amsmath,amssymb,
aps,
prb,
]{revtex4-1}

\usepackage{graphicx}% Include figure files
\usepackage{dcolumn}% Align table columns on decimal point
\usepackage{bm}% bold math
\begin{document}

\preprint{APS/123-QED}

\title{The effect of the frozen and pinned surface approximations on the spatial distribution of incompressible and compressible strips in quantum Hall regime}% Force line breaks with \\
%\thanks{A footnote to the article title}%

\author{Ahmet Emre KAVRUK}
\email{aekavruk@selcuk.edu.tr}
\author{Teoman \"{O}ZT\"{U}RK}%
 %\email{Second.Author@institution.edu}

%\collaboration{MUSO Collaboration}%\noaffiliation

\author{\"{U}lfet ATAV}
% \homepage{http://www.Second.institution.edu/~Charlie.Author}
%\affiliation{
% Second institution and/or address\\
% This line break forced% with \\
%}%
%\affiliation{
% Third institution, the second for Charlie Author
%}%
\author{H\"{u}seyin Y\"{U}KSEL}

\affiliation{%
 Selcuk University, Faculty of Science, 
Department of Physics, 42075 Konya,Turkey
}%

%\collaboration{CLEO Collaboration}%\noaffiliation

\date{\today}% It is always \today, today,
             %  but any date may be explicitly specified

\begin{abstract}
Pinned surface and frozen surface approximations are two commonly used approximations for the boundary conditions at the exposed surfaces of semiconductor structures.  We have studied the effect of pinned surface and frozen surface boundary conditions on the spatial distribution of compressible and incompressible strips observed in the two dimensional electron gas formed in a GaAs/AlGaAs heterostructure under quantum Hall effect regime. We have used semi classical Thomas-Fermi method for describing the many body problem along with the Poisson equation for electrostatics. We observe that the boundary conditions significantly effect the spatial distributions of the compressible and incompressible strips. 
\begin{description}
%\item[Usage]
%Secondary publications and information retrieval purposes.
\item[PACS numbers]{73.20.-r, 73.43.Cd, 71.70.Di, 02.70.-c}
%\pacs{73.20.-r}%\pacs{2}73.43.Cd\pacs{3}71.70.Di\pacs{4}02.70.-c
%{73.20.-r, 73.43.Cd, 71.70.Di, 02.70.-c}
%\pacs{73.20.-r, 73.43.Cd, 71.70.Di, 02.70.-c}
%May be entered using the \verb+\pacs{#1}+ command.
%\item[Structure]
%You may use the \texttt{description} environment to structure your abstract;
%use the optional argument of the \verb+\item+ command to give the category of each item. 
\end{description}
\end{abstract}

%\pacs{73.20.-r, 73.43.Cd, 71.70.Di, 02.70.-c}% PACS, the Physics and Astronomy
                             % Classification Scheme.
%\keywords{Suggested keywords}%Use showkeys class option if keyword
                              %display desired
\maketitle

%\tableofcontents

\section{\label{sec:level1}Introduction}

Quantum Hall Effect (QHE) has continuously been a field of attraction for both theoretical and experimental researchers since its discovery.\cite{Klitzing_1980} Theoretical works on this subject were concentrated on developing a basis to understand the changes in the longitudinal resistance and the formation of plateaus in the transverse (Hall) resistance.\cite{Chklovskii_1992,Chklovskii_1993,Lier_1994,Oh_1997,Guven_2003,Siddiki_Gerhardts_2003,Siddiki_Gerhardts_2004,Siddiki_2007,Siddiki_etal_2006,Siddiki_Marquardt_2007,Ihnatsenka_Zozoulenko_2008,Ihnatsenka_Zozoulenko_2006:73p075331,Ihnatsenka_Zozoulenko_2006:74p075320,Ihnatsenka_Zozoulenko_2006:74p201303R,Ihnatsenka_Zozoulenko_2006:73p155314,Ihnatsenka_Zozoulenko_2008:78p035340,Ihnatsenka_Zozoulenko_2007:75p035318} These works have led to the consensus that the plateaus in the transverse resistance and the changes in the longitudinal resistance can be understood in terms of compressible strips (CS) and incompressible strips (IS).  Therefore, determination of the spatial distribution of these strips is important for an accurate description of QHE.

In some experimental works these strips were used in place of electron or photon beams to construct devices equivalent to well known interferometer structures such as Mach-Zehnder or Aharonov-Bohm interferometers.\cite{Ji_2003,Neder_2006,Camino_2005} The spatial distribution of CS and IS is very crucial for understanding the behavior of these devices.\cite{Ihnatsenka_Zozoulenko_2008,Siddiki_Kavruk_2008,Ozturk_japon} In another experimental study it was observed that changing the magnetic field strength causes a hysteresis in the Hall resistance of bilayer systems with a density mismatch.\cite{Tutuc_2003,Pan_2005} It is believed that the underlying physics in this hysteresis phenomenon can also be understood in terms of the CS and IS.\cite{Siddiki_2007,Siddiki_etal_2006,Ihnatsenka_Zozoulenko_2007:75p035318,Kavruk_japon} The basic principles for determining spatial distribution of the CS and IS were laid by the seminal works of Chklovskiis.\cite{Chklovskii_1992,Chklovskii_1993}
 
CS and IS correspond to partially and fully occupied Landau levels which in turn are determined by the applied magnetic field. In a complete description, the charge distribution within the 2DEG should be determined from a solution of the Schroedinger equation and thermodynamic equilibrium. The effective potential term in the Schroedinger equation and the local thermodynamic equilibrium of the 2DEG are determined by the electrostatic potential. Therefore, the spatial distribution of CS and IS is basically determined by the electrostatics of the structure and a correct description of the electrostatics of the system is very important. It is well known that the electrostatics of a semiconductor structure is described by the Poisson Equation for dielectric media. The solution of the Poisson equation for a specific system is completely determined by the charge distribution and the boundary conditions of the system. Thus a correct description of the boundary conditions is very important. Naturally the surfaces of the semiconductor structure are the boundaries and it is clear that boundary condition for the electrostatic potential is the gate voltage where the surface is covered with gates. However, for the exposed parts of the surface the boundary condition is not clear. There exists two commonly used assumptions for the boundary conditions in the literature and they are called as Pinned Surface (PS) and Frozen surface (FS) boundary conditions.\cite{Laux_1988,Chen_1994,Davies_Larkin_1994,Davies_Larkin_1995,Larkin_Davies_1995}
 
For the PS boundary condition it is assumed that potential on exposed surfaces is fixed to the Fermi energy $E_F$ of the 2DEG.\cite{Davies_Larkin_1994,Davies_Larkin_1995,Larkin_Davies_1995} Under this assumption exposed surface is considered as a constant potential surface at zero potential, or in other words exposed surfaces are treated as grounded gates. In fact another plausible assumption for the PS boundary condition is that the potential on the exposed surface is fixed to the gate voltage $-V_g$ on the gates but in this case 2DEG cannot be confined to the region defined by the gates so that assuming the gates to be pinned to zero voltage is more acceptable. This assumption is more realistic at higher temperatures such as room temperature. Also, we would like to point that this boundary condition corresponds to an unrealistic sudden change of the potential at the gate edges from $–V_g$ to zero which is unrealistic. Because the value of the potential is fixed at the boundary PS approximation corresponds to a Dirichlet type boundary condition.\cite{Chen_1994}

FS approximation corresponds to the case where the charges on the surface are not affected from the gate potentials or the changes in the charge distribution in the rest of the system (i.e. the surface charges are frozen) and these surface charges induce a constant electric field in the normal direction of the surface. This approximation is expected to be more realistic at low temperatures. The temperatures required to observe QHE is around $ 1K$ so that the FS approximation seems more appropriate for an analysis of QHE. In the FS approximation the normal component of the electric field, i.e. the gradient of the electrostatic potential is fixed on the boundary and this type of boundary condition is called a Neumann boundary condition.\cite{Chen_1994} It is well known that numerical treatment of partial differential equations with Dirichlet boundary condition is easier than those with Neumann boundary conditions. This fact has led many researchers to use PS boundary conditions instead of FS boundary conditions.\cite{Siddiki_Marquardt_2007,Ihnatsenka_Zozoulenko_2008,Ihnatsenka_Zozoulenko_2006:73p075331,Ihnatsenka_Zozoulenko_2006:74p075320,Ihnatsenka_Zozoulenko_2006:74p201303R,Ihnatsenka_Zozoulenko_2006:73p155314,Ihnatsenka_Zozoulenko_2008:78p035340,Ihnatsenka_Zozoulenko_2007:75p035318,Siddiki_Kavruk_2008,Kavruk_japon,Ozturk_japon}

The effects of these boundary conditions on the charge distribution of 2DEG were compared for some semiconductor structures in the absence of magnetic field.\cite{Davies_Larkin_1994,Larkin_Davies_1995} These works have concentrated on the confinement potential on the 2DEG and formation of quantum wires or quantum dots. However there is no study comparing the effects of these approximations on the spatial distribution of the CS and IS under QHE regime. In this study, we will compare the effects of these boundary conditions on the spatial distribution of the IS and CS. In order to do so, we use Thomas-Fermi-Poisson (TFP) approach where quantum mechanical part of the problem is described by a semiclassical approach, namely Thomas-Fermi approximation and the electrostatic part of the problem is described by the Poisson equation. TFP approximation has extensively been used for the analysis of systems under QHE regime.\cite{Lier_1994,Oh_1997,Guven_2003,Siddiki_Gerhardts_2003,Siddiki_Gerhardts_2004,Siddiki_2007,Siddiki_etal_2006,Siddiki_Marquardt_2007}

\section{\label{sec:level2} Theory and The Model System}

\begin{figure}
\includegraphics[width=\columnwidth]{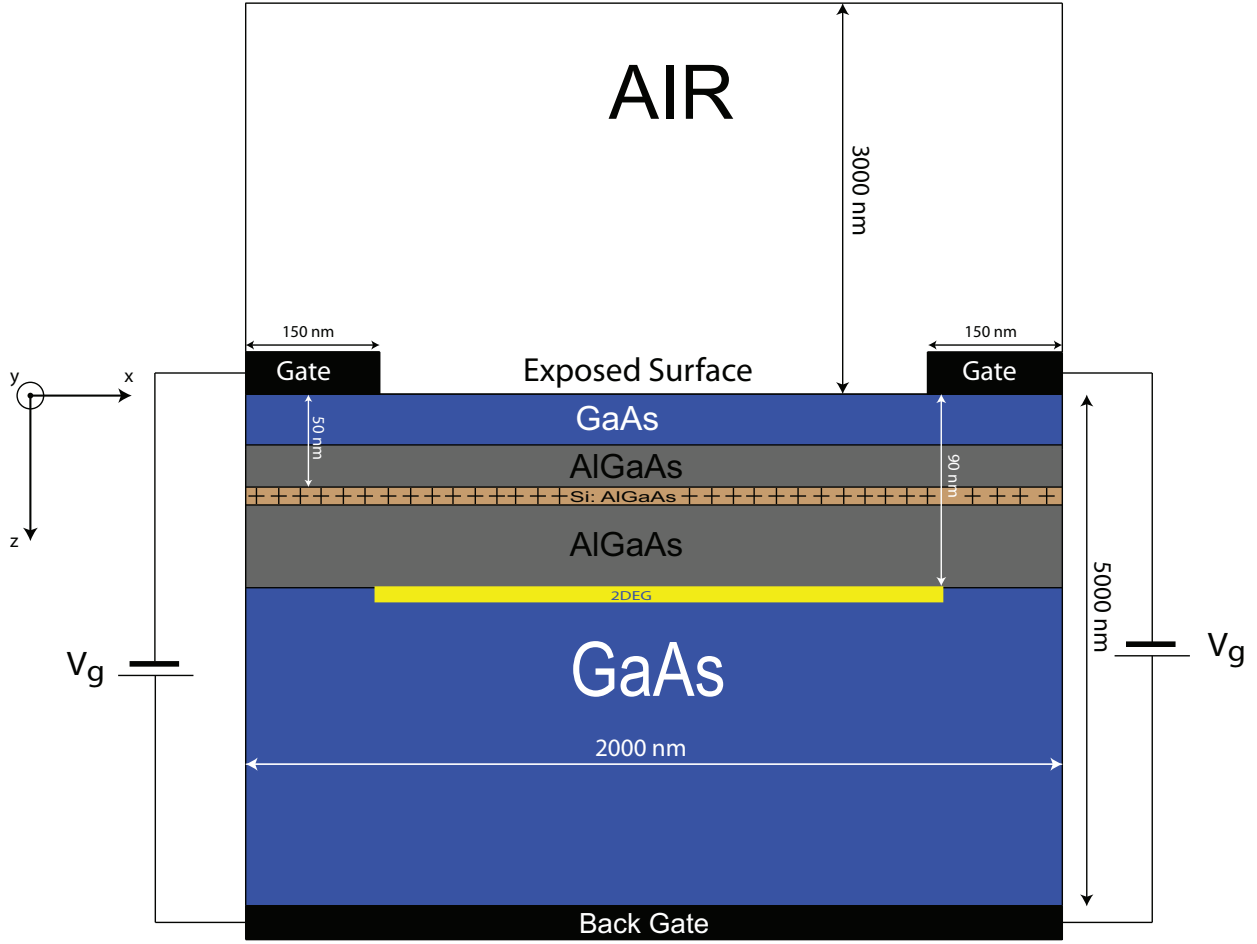}
\caption{A sketch of the model heterostructure used in the calculations.\label{model}}
\end{figure}

We consider the system shown in Fig.~\ref{model}. The system was supposed to be infinite and invariant in $ y$ direction, so that the problem was effectively reduced to a two dimensional problem. Also we have assumed that the structure is periodic in the $ x$ direction and we have used periodic boundary conditions in this direction. The $ z$ direction in Fig.~\ref{model} corresponds to the growth direction. We have assumed that the structure has a back gate at 0 V, i.e. the substrate is grounded. So that the boundary condition at the bottom is $V_{z=5\mu m}=0$. The boundary condition at the top surface is  $V_{z=0}=-V_g$ at the gates. Other parts of the top surface of the structure are exposed surface. We have performed separate calculations using both the FS and PS boundary conditions for this region.

Following Davies et al.\cite{Davies_Larkin_1994,Davies_Larkin_1995,Larkin_Davies_1995}, we have taken the electrostatic potential to be zero on the exposed surfaces under the PS approximation. The top surface separates the space into two domains and Poisson equation can be solved separately in each domain. However in this case we do not need the solution of the Poisson equation in the domain filled by air and solving Poisson equation for only the domain filled by the semiconductor structure in Fig.~\ref{model} is sufficient.

In the FS case we have assumed that the charges cannot move within the semiconductor structure so that the potential on any surface is not pinned to any constant value. In this case we treat the exposed surface as a free boundary and we assume that the gradient of the electrostatic potential is zero at some point sufficiently far from the surface of the structure. There are different layers within the semiconductor structure each one having different dielectric constants. However the differences between these dielectric constants are small and we have assumed that the dielectric constant is the same all over the semiconductor structure. However, considering the significant change in the dielectric constant on the exposed surface we impose continuity conditions on both the electrostatic potential and the dielectric displacement.

A sketch of the structure considered in this work is given in Fig.~\ref{model}. The depth of the semiconductor structure is taken as $5~\mu m$ along $ z$ direction. For the FS calculations, the healing distance in the air is taken as $3~\mu m$. The extent of the semiconductor structure along $x$ direction is $2~\mu m$. Each gate is assumed to be $150~nm$ wide, but due to the periodic boundary conditions along $ x$-direction the effective width of each gate is $300~nm$. We have assumed that $\delta$-doping is used in fabrication of the structure. The depth of the donors from the surface is taken as $50~nm$ while 2DEG is at $90~nm$ from the surface. The surface density of ionized donors and average electron density were assumed to be equal $n_{d}=\bar{n}_{e}=3.0\times 10^{15} m^{-2}$.

Encouraged by the slowly changing background potential, we have used the Thomas-Fermi (TF) approximation for the description of the quantum mechanical part of the problem. TF approximation presents a semi classical approach to obtain electron density in many-electron systems; it was used efficiently in many studies relevant to QHE because it provides simplicity in the calculations.\cite{Lier_1994,Oh_1997,Guven_2003,Siddiki_Gerhardts_2003} Within TF approximation the electron density is given as
\begin{equation} 
n_{el}(\vec{r})=\int dE D(E) f[(E+V(\vec{r})-\mu)/k_B T]
\end{equation}
where $D(E)$ is density of states, and $V(\vec{r})$ is the electrostatic potential, $\mu$ is the chemical potential, $k_B$ is Boltzmann constant and $T$ is the temperature while $f(\alpha)=(1+exp(\alpha))^{-1}$  is the well known Fermi distribution function. In the absence of magnetic field the density of states for 2DEG is constant and given as $D(E)=m/\pi\hbar^2$. On the other hand, under the influence of a magnetic field of strength B the density of states becomes $D(E)=\frac{g_s}{2\pi\ell^2}\sum_{n=0}^\infty \delta(E-E_n)$ here $g_s=2$ is spin degeneracy, $\ell=\sqrt{\hbar/eB}$ is magnetic length, $E_n=(n+1/2)\hbar\omega_c$ is energy of nth Landau level and $\omega_c=eB/m$ is cyclotron frequency. Therefore the electron density is given by
\begin{subequations}
\begin{equation}
n_{el}(\vec{r})=k_BT\frac{m}{\pi\hbar^2}\ln\left\lbrace 1+exp[(\mu-V(\vec{r}))/k_BT]\right\rbrace
\end{equation}
and
\begin{equation}
n_{el}(\vec{r})=\frac{g_s}{2\pi\ell^2}\sum_{n=0}^\infty f[(E_n+V(\vec{r})-\mu)/k_BT]
\end{equation}
\end{subequations}
in the absence and presence of a magnetic field, respectively.
 
The chemical potential $\mu$ appearing in Eqs.(2), is determined by the condition of conservation of the total number of electrons, that is
\begin{equation}
N_0=\int d^3\vec{r}~ n_{el}(\vec{r})
\end{equation}
here $N_0$ is total number of electrons. On the other hand the electrostatic potential must satisfy Poisson equation so that
\begin{equation}
\vec{\nabla}\ldotp(\varepsilon\vec{\nabla}V(\vec{r}))=-\rho(\vec{r})
\end{equation}
where $\rho(\vec{r})$ is total charge density and $\varepsilon$ is the dielectric constant of the material. Total charge density can be written as $\rho(\vec{r})=\rho_d(\vec{r})-en_{el}(V\vec{r})$ where $\rho_d(\vec{r})$ is the fixed charge density due to donors and impurities except 2DEG. Equations (2-4) form a closed set of coupled equations. Electrostatic potential, chemical potential and electron distribution in the 2DEG should be obtained from a simultaneous solution of these equations.  However, during the simultaneous solutions of these equations one may encounter many divergence and instability problems. In order to overcome these problems we have used a slightly modified version of an iterative algorithm previously used by Gerhardts and co-workers.\cite{Lier_1994,Oh_1997,Guven_2003,Siddiki_Gerhardts_2003} This algorithm can be summarized as follows:

First, in order to create an initial value for the iterative steps, Eqs.(2-4) are solved self-consistently and electron distribution is obtained in the absence of magnetic field at absolute zero temperature. Then using this first solution as an initial value at a sufficiently high temperature T in the absence of magnetic field Eqs.(2-4) are solved again self-consistently and chemical potential $\mu$, electrostatic potential $V(\vec{r})$ and electron density $n_{el}(\vec{r})$ are obtained. We use this finite temperature solution as an initial value to obtain a solution for the Eqs.(2-4)  for the same temperature but now in the presence of the magnetic field. Afterwards the temperature is gradually reduced until we reach the target temperature. For the calculations presented in this study, we have chosen the initial high temperature and the target temperature as $60~K$ and $1.4~K$, respectively. While decreasing the temperature, the result obtained in the previous step is used as the initial value at each step. For the iterative solution, which forms the basic step of this algorithm, we used a generalized form of Newton-Raphson Method details of which is described below.

Let $\mu^*$ and $V^*(\vec{r})$ be the chemical potential and the electrostatic potential corresponding to the exact simultaneous solution of Eqs. (2-4). On the other hand we use the potential function $V(\vec{r})$ obtained in the previous step as a first estimate. If we indicate the difference between the exact and estimated potentials as $\delta V(\vec{r})$, then we can write the exact potential as
\begin{equation}
V^*(\vec{r})=V(\vec{r})+\delta V(\vec{r}).
\end{equation}
The exact potential satisfies Poisson equation, i.e.
\begin{equation}
\nabla^2V^*=\nabla^2V+\nabla^2\delta V=-\frac{1}{\varepsilon}(\rho_d-e n_{el}(V+\delta V)).
\end{equation}
If we expand $n_{el}$ in a Taylor series about $V$ and keep only the first order terms
\begin{equation}
\nabla^2V+\nabla^2\delta V=-\frac{1}{\varepsilon}(\rho_d-e n_{el}(V+\delta V))+\frac{e}{\varepsilon}\frac{\partial n_{el}}{\partial V}\delta V
\end{equation}
is obtained. According to Eqs. (2), we can write the relation between derivatives of electron density $n_{el}$ with respect to electrostatic and chemical potentials as
\begin{equation}
\frac{\partial n_{el}}{\partial V}=-\frac{\partial n_{el}}{\partial \mu}.
\end{equation}
So Eq. (7) can be written as
\begin{equation}
\nabla^2V+\nabla^2\delta V=-\frac{1}{\varepsilon}(\rho_d-e n_{el}(V+\delta V))-\frac{e}{\varepsilon}\frac{\partial n_{el}}{\partial \mu}\delta V
\end{equation}
or
\begin{equation}
\nabla^2\delta V+\frac{e}{\varepsilon}\frac{\partial n_{el}}{\partial \mu}\delta V=-\nabla^2V-\frac{1}{\varepsilon}(\rho_d-e n_{el}(V+\delta V))
\end{equation}
Here if we define the functions $p=\frac{e}{\varepsilon}\frac{\partial n_{el}}{\partial \mu}$ and $f=-\nabla^2V-\frac{1}{\varepsilon}(\rho_d-e n_{el}(V+\delta V))$ then Eq. (10) reduces to
\begin{equation}
\nabla^2\delta V+p\delta V=f.
\end{equation}
Equation (11) is a second order differential equation for the correction term $\delta V$ however it involves a linear term unlike the Poisson equation. By adding $\delta V$, which is obtained from the solution of Eq. (11), to the previous potential estimate we obtain a better estimate for the electrostatic potential function. If we denote the potential estimates obtained in two successive steps as $V^{n+1}$ and $V^n$ then one step of the Newton-Raphson algorithm can be expressed as
\begin{equation}
V^{n+1}=V^{n}+\delta V.
\end{equation}
Then the chemical potential and electron density which correspond to this new potential are obtained by using Eqs. (2) and Eq. (3). This procedure is repeated until the changes between two successive steps in the potential, electron density and chemical potential are small enough. 

\section{\label{sec:level3} Results and Discussion}

It is already pointed out that the boundary conditions are very important in obtaining the solution of Poisson equation. These boundary conditions reflect the structure and the behavior of the surface and have important effects on the behavior of 2DEG. In fact the purpose of this study is to investigate how different boundary conditions affect the behavior of 2DEG under intense magnetic field. Therefore we have performed calculations for physically plausible two different boundary conditions for the upper (exposed) surface of the structure. These boundary conditions correspond to two different surface behaviors, namely PS and FS approximations. We have used the model structure shown in Fig.~\ref{model} in all our calculations. We have chosen the gate voltages so that the potential difference between the gates and the center of 2DEG is $ -0.35~V$.

\begin{figure}
\includegraphics[width=\columnwidth]{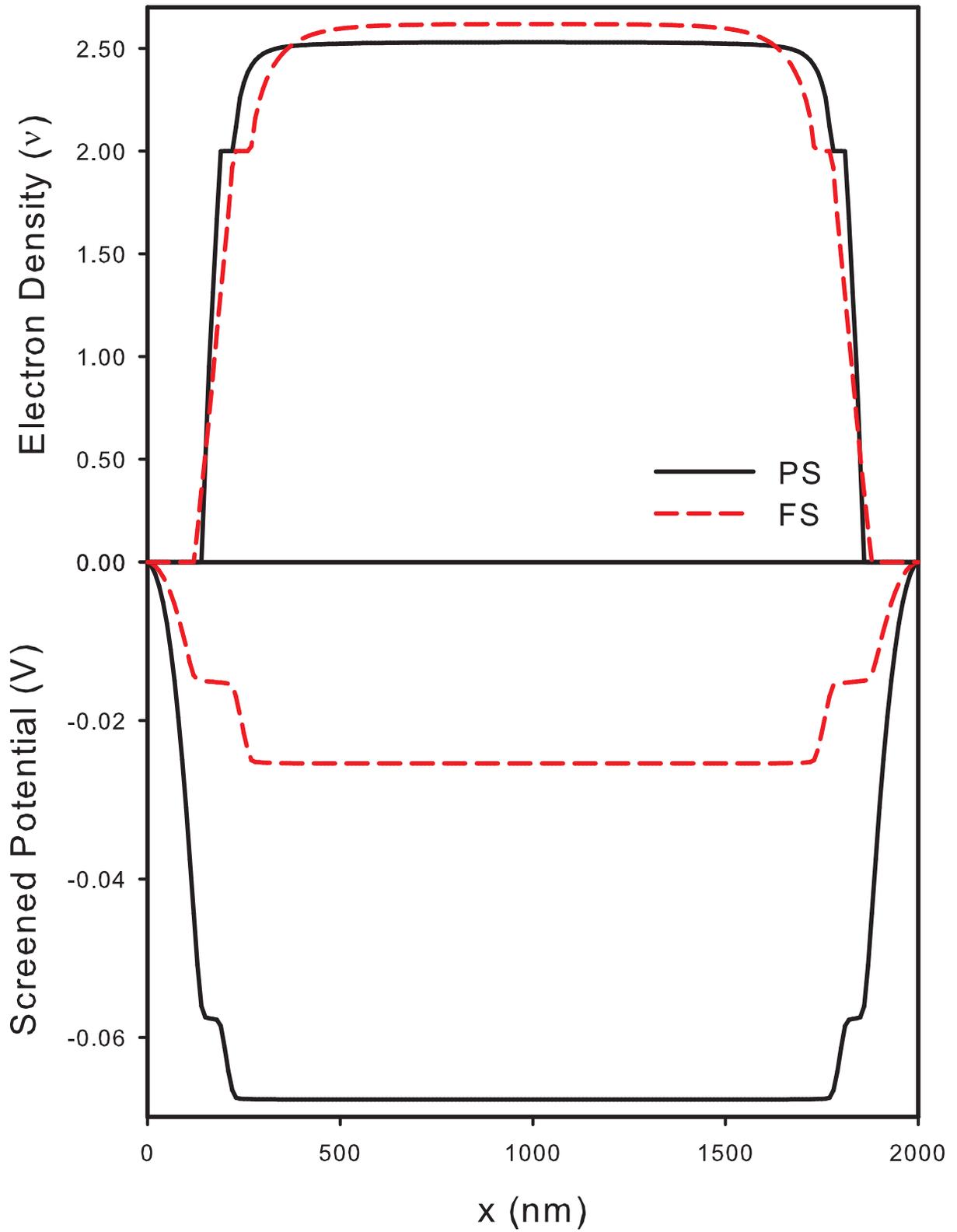}
\caption{Variation of the electron density and screened potential in the 2DEG for $B=6.0~ T$.The solid(black) lines correspond to the PS and the dashed(red) lines correspond to the FS boundary conditions. \label{6.0T}}
\end{figure}

We present our results in Fig.~\ref{6.0T} for electron density distribution and variation of the screened potential under both FS and PS boundary conditions for a magnetic field intensity of $6.0~T$. The solid(black) lines correspond to the PS and the dashed(red) lines correspond to the FS boundary conditions. In order to compare the potentials we have given the potential difference from the edge of the structure in the plane of 2DEG.  Under the PS boundary conditions, the potential at the top surface of the structure is forced to an unphysical jump at gate edges, even though this sharp change smooth as we move away from the surface its effect still can be seen in the plane of 2DEG. In contrast to the PS approximation, the potential in the FS approximation is not forced to a jump at the exposed surface, and thus the potential profile on the plane of 2DEG also exhibits a more smooth behavior. The sharp change in the potential results in a wider depletion region (i.e. a region without any electrons) at the edges and rises more quickly than those obtained under FS approximation. The smoother change in the potential under FS approximation results in a slower change in the electron density and under this approximation electrons penetrate more toward the edge of the structure so that the depletion region is narrower. While the electron density changes more rapidly near the edges under PS approximation, it quickly saturates and behaves like constant near the center. This quick saturation should be expected because the distance between the exposed surface and 2DEG is very small and the potential is assumed to be constant at the exposed surface. This behavior of the electron density changes completely for FS approximation where the electron density and the electrostatic potential changes more slowly.

\begin{figure}
\includegraphics[width=\columnwidth]{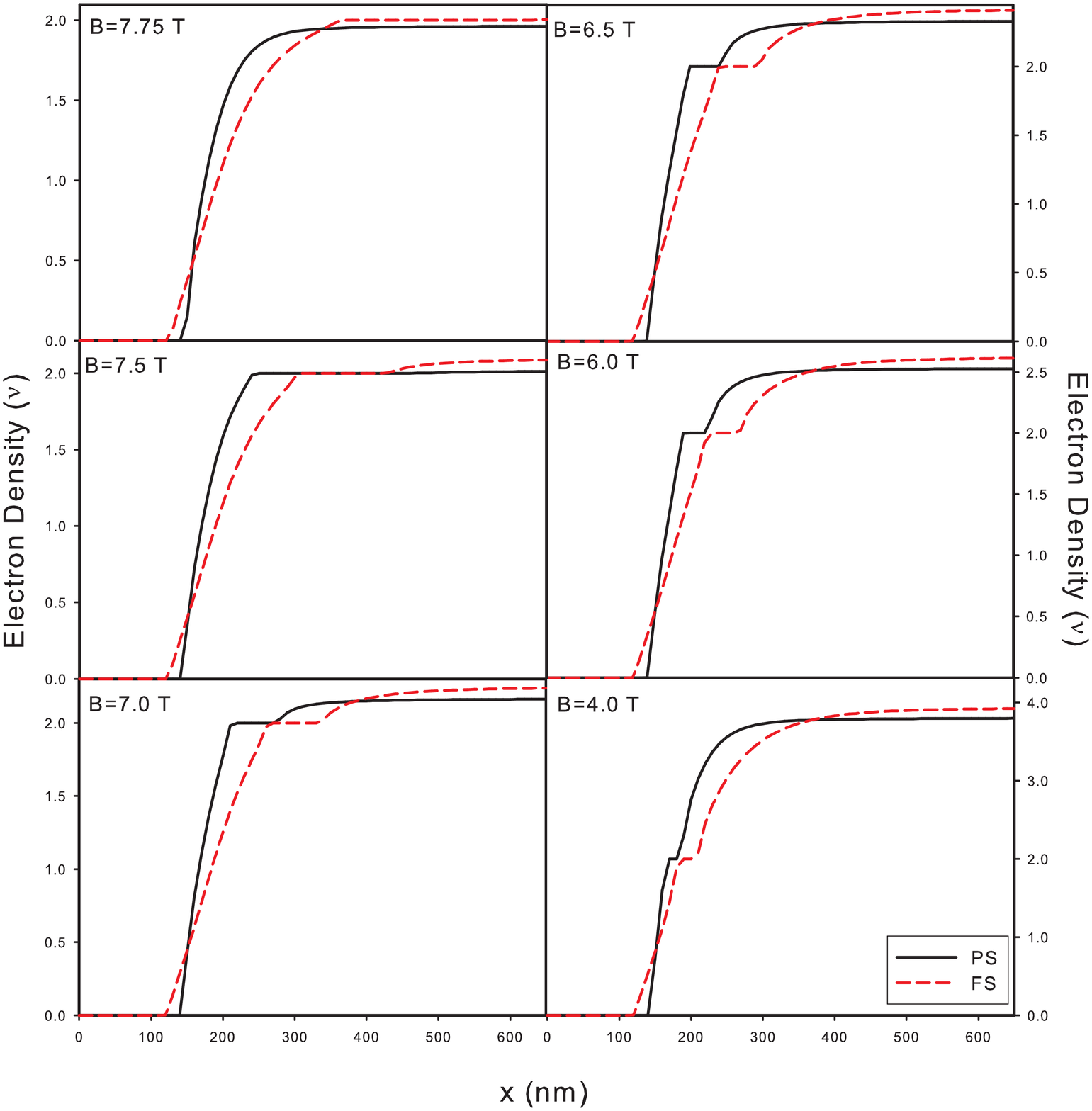}
\caption{Spatial electron density distributions in terms of filling factor for several magnetic fields.The solid(black) lines correspond to the PS and the dashed(red) lines correspond to the FS boundary conditions. \label{ner}}
\end{figure}

\begin{figure}
\includegraphics[width=\columnwidth]{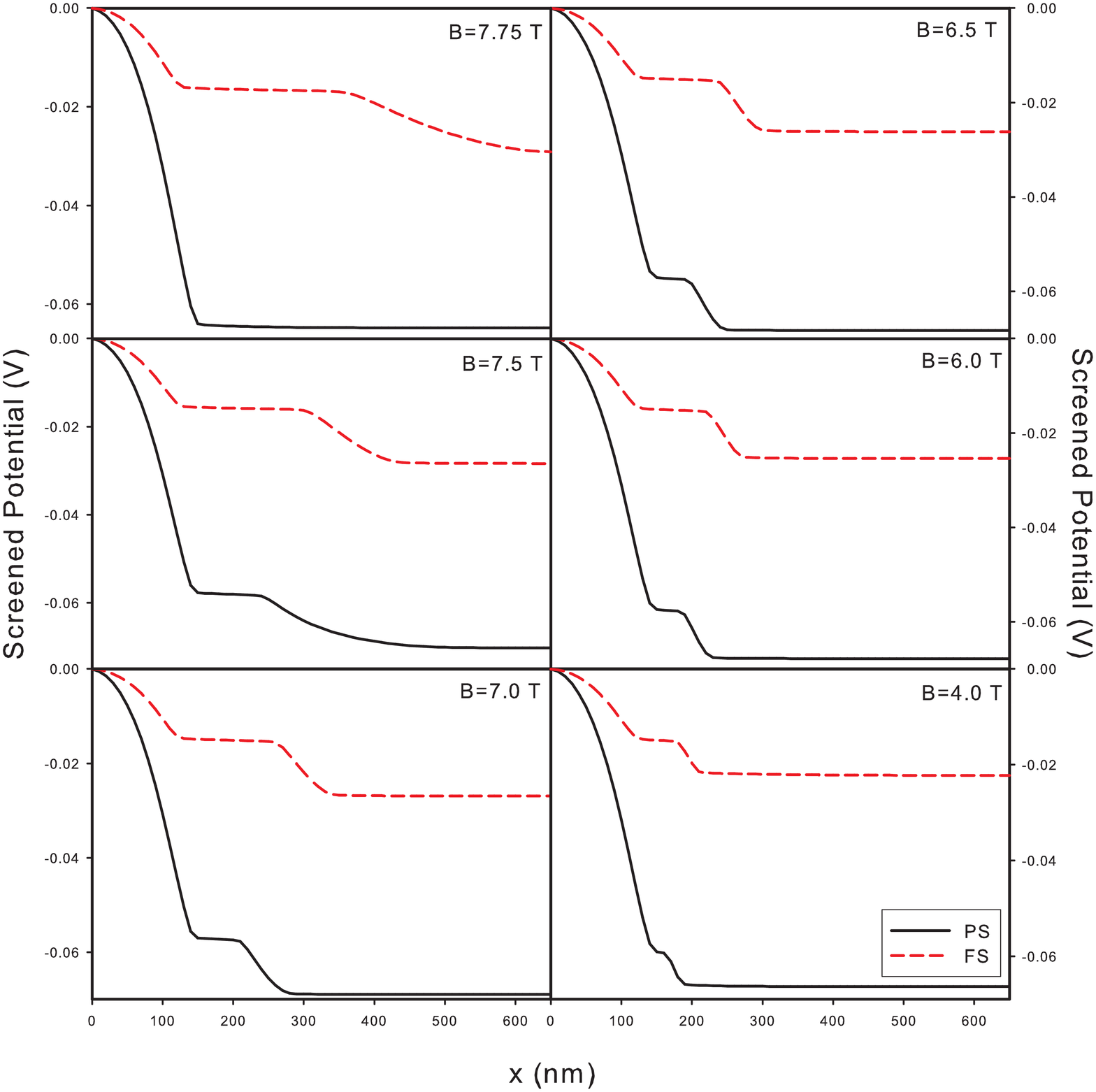}
\caption{Spatial distribution of the screened potential for several magnetic fields.The solid(black) lines correspond to the PS and the dashed(red) lines correspond to the FS boundary conditions. \label{pot}}
\end{figure}

Because the structure is symmetrical with respect to the center and the changes in the potential and the electron density occurs near to the edges we plot the rest of our results for a smaller range at the left edge of the structure. In Fig.~\ref{ner} we present the electron densities obtained for various magnetic field strengths. The potential profiles for the same magnetic field strengths are given in Fig.~\ref{pot}. It can be seen from Fig.~\ref{ner} that IS move toward edge of the structure and get narrower with decreasing magnetic field. Hall plateaus in the system occur if there is at least one IS in the system\cite{Ahlswede_2001,Ahlswede_2002} and when the width of the IS is less then cyclotron radius of an electron, the system leaves the plateau.\cite{Siddiki_Gerhardts_2004} When one uses the PS approximation the first appearance of IS occurs at lower magnetic field strengths. In Fig.~\ref{ner}, there is no IS observed for $B=7.75~T$ and IS appears for $ B=7.5~T$ under PS approximation while under FS approximation IS is observed for both $ B=7.75~T$ and $ B=7.5~T$. Also the sharp change of the potential under PS approximation results in even narrower IS. Thus under PS approximation one predicts the system to leave plateau region at a higher magnetic field strength.  As a result, using PS approximation may lead to wrong predictions of narrower plateau regions.

The total electrostatic potential is called as screened potential when it is obtained self consistently because it also contains the screening effect of the 2DEG. Screened potential profiles obtained for various magnetic field strengths are given in Fig.~\ref{pot}. Because in the CS region 2DEG behaves like a metal the potential is perfectly screened and behaves like constant. Whereas in the IS regions the potential cannot be screened and changes until it reaches to the next Landau level. So the screened potential is constant while the electron density changes and electron density stays constant while the potential changes. However, the effect of the two approximations for the boundary conditions at the exposed surface on screened potential profiles is similar to the electron density.

It is generally accepted that FS boundary conditions reflect the real physical conditions better for semiconductor structures at low temperatures.\cite{Laux_1988,Davies_Larkin_1995,Larkin_Davies_1995} However, one has to consider the region beyond the exposed surface in some way under FS boundary conditions. When working with FS boundary conditions a complete description of the problem requires the solution of Poisson equation for each electron distribution in the 2DEG. On the other hand, working with PS boundary conditions is much easier. The exposed surface under PS boundary conditions is treated as a grounded gate and the mirror charge convention may be used for a complete description of the problem. Then solving the Poisson equation for the gates and the donor distribution only once will be sufficient. In conclusion, we observe that using PS approximation for exposed surface boundary conditions leads one to obtain erroneous results for the spatial distribution of IS and CS under QHE regime. However most of the studies in the literature refrain from using FS boundary conditions because it is easier to work with PS.\cite{Siddiki_Marquardt_2007,Ihnatsenka_Zozoulenko_2008,Ihnatsenka_Zozoulenko_2006:73p075331,Ihnatsenka_Zozoulenko_2006:74p075320,Ihnatsenka_Zozoulenko_2006:74p201303R,Ihnatsenka_Zozoulenko_2006:73p155314,Ihnatsenka_Zozoulenko_2008:78p035340,Ihnatsenka_Zozoulenko_2007:75p035318,Siddiki_Kavruk_2008,Kavruk_japon,Ozturk_japon}

\begin{acknowledgments}
This work is supported by Selcuk University BAP Grant No. 07101037.
\end{acknowledgments}

%\bibliography{apssamp}% Produces the bibliography via BibTeX.

\begin{thebibliography}{00}

\bibitem{Klitzing_1980} K. v. Klitzing, G. Dorda and M. Pepper, Phys. Rev. Lett. {\bf 45,} 494 (1980).
\bibitem{Chklovskii_1992} D. B. Chklovskii, B. I. Shklovskii and L. I. Glazman, Phys. Rev. B {\bf 46,} 4026 (1992).
\bibitem{Chklovskii_1993} D. B. Chklovskii, K. A. Matveev and B. I. Shklovskii, Phys. Rev. B {\bf 47,} 12605 (1993).
\bibitem{Lier_1994} K. Lier and R. R. Gerhardts, Phys. Rev. B {\bf 50,} 7757 (1994).
\bibitem{Oh_1997} J. H. Oh and R. R. Gerhardts, Phys. Rev. B {\bf 56,} 13519 (1997).
\bibitem{Guven_2003} K. Guven and R. R. Gerhardts, Phys. Rev. B {\bf 67,} 115327 (2003).
\bibitem{Siddiki_Gerhardts_2003} A. Siddiki and R. R. Gerhardts, Phys. Rev. B {\bf 68,} 125315 (2003).
\bibitem{Siddiki_Gerhardts_2004} A. Siddiki and R. R. Gerhardts, Phys. Rev. B {\bf 70,} 195335 (2004).
\bibitem{Siddiki_2007} A. Siddiki, Phys. Rev. B {\bf 75,} 155311 (2007).
\bibitem{Siddiki_etal_2006} A. Siddiki, S. Kraus and R. R. Gerhardts, Physica E {\bf 34,} 136 (2006).
\bibitem{Siddiki_Marquardt_2007} A. Siddiki and F. Marquardt, Phys. Rev. B {\bf 75,} 045325 (2007).
\bibitem{Ihnatsenka_Zozoulenko_2008} S. Ihnatsenka and I. V. Zozoulenko, Phys. Rev. B {\bf 77,} 235304 (2008).
\bibitem{Ihnatsenka_Zozoulenko_2006:73p075331} S. Ihnatsenka and I. V. Zozoulenko, Phys. Rev. B {\bf 73,} 075331 (2006).
\bibitem{Ihnatsenka_Zozoulenko_2006:74p075320} S. Ihnatsenka and I. V. Zozoulenko, Phys. Rev. B {\bf 74,} 075320 (2006).
\bibitem{Ihnatsenka_Zozoulenko_2006:74p201303R} S. Ihnatsenka and I. V. Zozoulenko, Phys. Rev. B {\bf 74,} 201303(R) (2006).
\bibitem{Ihnatsenka_Zozoulenko_2006:73p155314} S. Ihnatsenka and I. V. Zozoulenko, Phys. Rev. B {\bf 73,} 155314 (2006).
\bibitem{Ihnatsenka_Zozoulenko_2008:78p035340} S. Ihnatsenka and I. V. Zozoulenko, Phys. Rev. B {\bf 78,} 035340 (2008).
\bibitem{Ihnatsenka_Zozoulenko_2007:75p035318} S. Ihnatsenka and I. V. Zozoulenko, Phys. Rev. B {\bf 75,} 035318 (2007).
\bibitem{Ji_2003} Y. Ji, Y. Chung, D. Sprinzak, M. Heiblum, D. Mahalu and H. Shtrikman, Nature {\bf 422,} 415 (2003).
\bibitem{Neder_2006} I. Neder, M. Heiblum, Y. Levinson, D. Mahalu and V. Umansky, Phys. Rev. Lett. {\bf 96,} 016804 (2006).
\bibitem{Camino_2005} F. E. Camino, W. Zhou and V. J. Goldman, Phys. Rev. B {\bf 72,} 155313 (2005).
\bibitem{Siddiki_Kavruk_2008} A. Siddiki, A. E. Kavruk, T. Ozturk, U. Atav, M. Sahin and T. Hakioglu, Physica E {\bf 40,} 1398 (2008).
\bibitem{Ozturk_japon} T. Ozturk, A. E. Kavruk, A. Ozturk, U. Atav and H. Yuksel, Journal of Physics:Conference Series {\bf 334,} 012034 (2011).
\bibitem{Tutuc_2003} E. Tutuc, R. Pillarisetty, S. Melinte, E. P. De Poortere, and M. Shayegan, Phys. Rev. B {\bf 68} 201308(R) (2003).
\bibitem{Pan_2005} W. Pan, J. L. Reno, and J. A. Simmons, Phys. Rev. B {\bf 71} 153307 (2005).
\bibitem{Kavruk_japon} A. E. Kavruk, T. Ozturk, A. Ozturk, U. Atav and H. Yuksel, Journal of Physics:Conference Series {\bf 334} 012066 (2011).
\bibitem{Laux_1988} S. E. Laux, D. J. Frank and F. Stern, Surface Science {\bf 196,} 101 (1988).
\bibitem{Chen_1994} M. Chen, W. Porod and D. J. Kirkner, J. Appl. Phys. {\bf 75,} 2545 (1994).
\bibitem{Davies_Larkin_1994} J. H. Davies and I. A. Larkin, Phys. Rev. B {\bf 49,} 4800 (1994).
\bibitem{Davies_Larkin_1995} J. H. Davies, I. A. Larkin and E. V. Sukhorukov, J. Appl. Phys. {\bf 77,} 4504 (1995).
\bibitem{Larkin_Davies_1995} I. A. Larkin and J. H. Davies, Phys. Rev. B {\bf 52,} R5535 (1995).
\bibitem{Ahlswede_2001} E. Ahlswede, P. Weitz, J. Weis, K. v. Klitzing and K. Eberl, Physica B {\bf 298,} 562 (2001).
\bibitem{Ahlswede_2002} E. Ahlswede, J. Weis, K. v. Klitzing and K. Eberl, Physica E {\bf 12,} 165 (2002).
\end{thebibliography}

\end{document}